# On the distribution of DNA translocation times in solid-state nanopores: an analysis using Schrödinger's first-passage-time theory

**Daniel Y. Ling[1] and Xinsheng Sean Ling[2*]**

[1]Princeton University, Princeton, NJ 08544
[2]Department of Physics, Brown University, Providence, RI 02912

**Abstract**

In this short note, a correction is made to the recently proposed solution [1] to a 1D biased diffusion model for linear DNA translocation and a new analysis will be given to the data in [1]. It was pointed out [2] by us recently that this 1D linear translocation model is equivalent to the one that was considered by Schrödinger [3] for the Enrenhaft-Millikan measurements [4,5] on electron charge. Here we apply Schrödinger's first-passage-time distribution formula to the data set in [1]. It is found that Schrödinger's formula can be used to describe the time distribution of DNA translocation in solid-state nanopores. These fittings yield two useful parameters: drift velocity of DNA translocation and diffusion constant of DNA inside the nanopore. The results suggest two regimes of DNA translocation: (I) at low voltages, there are clear deviations from Smoluchowski's linear law of electrophoresis [6] which we attribute to the entropic barrier effects; (II) at high voltages, the translocation velocity is a linear function of the applied electric field. In regime II, the apparent diffusion constant exhibits a quadratic dependence on applied electric field, suggesting a mechanism of Taylor dispersion effect likely due the electro-osmotic flow field in the nanopore channel. This analysis yields a dispersion-free diffusion constant value of 11.2 nm$^2$/μs for the segment of DNA inside the nanopore which is in agreement with Stokes-Einstein theory quantitatively. The implication of Schrödinger's formula for DNA sequencing is discussed.

**Introduction**

Ever since the first demonstration of electric-field driven linear "translocation" (i.e. transport) of long nucleic acid molecules through a nanopore [7], there have been intense interests in searching for an analytical form for analyzing the time distributions observed in a wide range of systems, ssDNA [8] in α-hemolysin, dsDNA [9,10] and proteins [11,12] in solid-state nanopores. Such analyses could provide useful insights into the nature of DNA translocation dynamics in nanopores. Ultimately the usefulness of a nanopore technology will depend on the correct interpretation of the translocation signals [13].

Even before the experimental observations of electric-field driven DNA translocation, Sung and Park [14] theoretically considered the DNA translocation driven by a chemical potential gradient. They predicted that the mean first-passage time should scale as a power law of the DNA length due to the configurational entropy effects. Subsequently, Lubensky and Nelson [15] proposed a 1D biased diffusion model for DNA translocation in which the DNA is modeled as a rigid rod and the configuration entropy of the DNA is ignored. They derived an analytical form for the first-passage probability distribution function (FP-PDF), which is the first-passage-time distribution function measured in experiments, using absorbing boundary conditions on both the *cis* and *trans* sides.

Most recently, a 1D biased diffusion model for DNA translocation has been proposed [1] with only one absorbing boundary on the *trans* side. Unfortunately a mathematical error was made leading to the solutions for the Fokker-Planck equation, compromising the validity of the FP-PDF formula, as pointed out previously [2]. Here we summarize the key points of [2] in a paper form and apply the correct form of FP-PDF to the DNA translocation data in [1].

**Models of DNA translocation: the first-passage problem in 1D biased diffusion**

In spite of the intense theoretical effort in the past decade, many aspects of electric-field driven DNA translocation are still not well understood. First, the DNA capture into the pore is a process in which the DNA loses some of its configurations. Sung and Park [14], Muthukumar [16], and others have considered the effects of an entropic barrier for DNA entry into the *cis* side of the pore. The entropic barrier concept provides a good explanation for the observed behavior in the increased capturing rate in experiments in α-hemolysin [8] and solid-state nanopores [17]. Secondly, the translocation times, the duration of a current blockage event, have been found to exhibit a wide distribution.

The first attempt at constructing an analytical expression for the distribution of the translocation times was proposed by Lubensky and Nelson [15], motivated by the experiment of Kasianowicz *et al*.[7] In the Lubensky-Nelson model, the DNA translocation is modeled as a 1D biased random walk, effectively treating the pore as a random walker diffusing through the contour length of the DNA. In the Lubensky-Nelson model, the fluctuations of the DNA tails on the *cis* side and the heads in the *trans* side may contribute to the effective diffusion constant and the drag coefficient, they do not appear explicitly in the Fokker-Planck equation. They then proceeded to solve the corresponding Fokker-Planck equation with absorbing boundary conditions at both *cis* and *trans* sides. They derived an explicit probability density function (PDF) for the first-passage times. Later, Berezhkovskii and Gopich [18] argued that the absorbing boundary condition on the *cis* side in the Lubensky-Nelson model as being unphysical, and they offered alternative "radiation" boundary conditions with which they produced a numerical solution to FP-PDF.

**Schrödinger's first-passage time theory:**

Most recently, Li and Talaga [1] further simplified the 1-D biased diffusion model for DNA translocation by assuming only one absorbing boundary condition for the *trans* side. On the *cis* side, the probability density function $P(x, t)$ is only constrained by the initial condition $P(x, 0) = \delta(x)$. Unfortunately a mathematical error was made in deriving the FP-PDF. The first objective of this note is to correct this error [1]. As pointed in [2] that the model proposed by Li and Talaga (with one absorbing boundary condition) was in fact mathematically equivalent to the one that was solved by Schrödinger for the Enrenhaft-Millikan experiments. We show here that the Schrödinger's FP-PDF can indeed be used for analyzing DNA translocation time distribution. For that matter, we find that the formula given by Lubensky and Nelson, as well as by Li and Talaga, can give equally good fittings to the translocation time distributions. However, only Schrödinger's FP-PDF allows us to extract a drift velocity that is consistent with Smoluchowski' law of linear electrophoresis.

In typical DNA translocation experiment, the membrane thickness is significantly shorter than the contour length L of the DNA. Lubensky and Nelson argued that the translocation process of the DNA going through the pore can be viewed, equivalently, as the pore undergoing 1-D biased Brownian motion along the DNA. The distribution of the measured translocation times should be that of the probability density function of the first-passage times of 1-D biased diffusion. In this model, the probability density function (PDF) $P(x, t)$ (per unit length) of finding the pore on the position $x$ on the DNA can be obtained by solving the Fokker-Planck equation with proper boundary conditions.

$$\frac{\partial P(x,t)}{\partial t} = D\frac{\partial^2 P(x,t)}{\partial x^2} - v\frac{\partial P(x,t)}{\partial x} \qquad (1)$$

where D and v are the diffusion constant and the drift velocity of the segment of the DNA inside the pore, respectively.

The first step of solving a first-passage problem is to solve the Fokker-Planck equation (1). For a differential equation such as (1), it is clear that the solution is critically dependent on the initial and boundary conditions. Here we will not repeat the arguments of Lubensky-Nelson [15] and Berezhkovskii-Gopich [18] for their choices of boundary conditions. For simplicity, we adopt the same approach as that of Li and Talaga [1] in assuming the initial condition of DNA capture into the pore as:

$$P(x, 0) = \delta(x) \qquad (2)$$

and the absorbing boundary once the DNA has completely translocated into the *trans* side of the pore:

$$P(L, t) = 0. \qquad (3)$$

Physically, the absorbing boundary condition at x = L reflects the physics that the translocated DNA cannot travel back into the pore. By not imposing a boundary condition for $P(x,t)$ at x = 0 is equivalent to allowing the DNA to retract to the *cis* side, which can indeed occur at least for DNA translocation in solid-state nanopores. (In the case of Schrödinger's work [3], this condition allows the Brownian walker to move back into the x < 0 part of the space. For the Enrenhaft-Millikan experiments, this is physically reasonable since x=0 is arbitrarily set in measurement window.)

A practical advantage of not imposing a boundary condition at the *cis* side is that the problem becomes exactly solvable. In fact, the problem becomes equivalent to the 1D biased diffusion model solved by Schrödinger. With these conditions, the solution to the Fokker-Planck equation is as follows:

$$P(x,t) = \frac{1}{\sqrt{4\pi Dt}} (e^{-(x-vt)^2/4Dt} - Ae^{-(x-2L-vt)^2/4Dt}) \quad (4)$$

where $A = exp(vL/D)$. One can check that this probability density function satisfies eq. (1) and the initial and boundary conditions (2) and (3). This solution was first given [3] by Schrödinger in 1915 to describe the Enrenhaft-Millikan experiments [4,5]. (In [1], to make $P(x,t)$ satisfy the boundary condition, $A=1$ and a positive sign for the $vt$ term in the argument of the second exponential were chosen. It is straightforward to show that such a form for $P(x, t)$ does not satisfy the Fokker-Planck equation in Eq. (1).)

The first passage probability density function (FP-PDF) (per unit time) is defined [3] as

$$F_1(t) = -\frac{d}{dt} \int_{-\infty}^{L} P(x,t) dx \quad (5)$$

and it has the physical meaning of being the probability per unit time for the random walker to pass the absorbing edge at x = L, i.e. the DNA has fully translocated through the pore. The result for $F_1(t)$ can be shown (see [19,20]) to be:

$$F_1(t) = \frac{L}{\sqrt{4\pi Dt^3}} e^{-(L-vt)^2/4Dt} \quad (6)$$

By definition of a probability density function, $F_1(t)$ needs to be normalized, *i.e.* $\int_0^\infty F_1(t) dt = 1$. One can show that Eq. (6) is indeed normalized. (In contrast, the expression given by Li and Talaga [1] for $F_1(t)$ contains an extra term of $vt$ in addition to $L$ in the numerator, an error propagated from the choice of parameters in Eq. (4). One can show that this extra term will make $F_1(t)$ nonphysical since its integral over time is equal to 2.)

By assuming only one absorbing condition at the trans side, $x = L$, the solutions in $P(x,t)$ in equation (4) and in $F_1(t)$ in Eq. (5) contain the trajectories of the random walker travelled to $x < 0$ side and come back to the region $0 < x < L$. During DNA translocation, at low voltages, there is finite possibility that the DNA can retract into the *cis* side against the electric field. At low voltages, the entropic effects are expected to play a role as the cost in free energy (loss of configurational entropy) can exert a retracting force. For pure 1D random walk, the travelling back into the x < 0 space is caused by thermal forces from the heat bath alone.

At high bias voltage, the DNA retraction after being captured should be rare. Thus we expect the solution in Eq. (6) to be of some guidance for data analysis in DNA translocation experiments. In what follows, we will use the Schrödinger FP-PDF in Eq. (6) to re-analyze the DNA translocation data in solid-state nanopores previously published in [1].

**Re-analysis of translocation data in [1]:**

Figure 1 shows the translocation time distributions at 4 selected voltages V = 20, 50, 80, and 110 mV provided by Prof. Jiali Li [1]. The experimental details were discussed in [1]. Briefly, the experiment was conducted on 4 kbp dsDNA in 1.6 M KCl with 20% glycerol at pH=7.5 in a nanopore (dia.~ 8 nm), housed in a 20 nm thick freestanding silicon nitride ($Si_3N_4$) membrane [1].

The red lines in Fig.1 are best fits to Eq. (6). Eq. (6) offers excellent description for all of the distribution data in [1]. However, we should point out that equally good fittings can be achieved [21] using the formula derived by Lubensky and Nelson [15] as well as the incorrect FP-PDF formula from [1]. A more rigorous test will be based on whether the fitting parameters make physical sense. Indeed, the extracted parameters using the Lubensky-Nelson do not show the expected dependence of drift velocity on applied bias voltages [2].

The extracted values for drift velocity and diffusion constant are shown in Figure 2 below. In Figure 2(A), one can clearly identify two regimes of DNA translocations. In regime I, at low voltages, the extracted drift velocity clearly deviates from the expected Smoluchowski's linear electrophoresis, Eq. (7) below. In contrast, in regime II at high voltages, this linear dependence is obeyed.

According to Smoluchowski [6], a charged object such as DNA in salty buffer conditions will undergo drift under the influence of applied electric field $E$. The drift velocity,

$$v = \mu E \qquad (7)$$

where $\mu$ is the electrophoretic mobility. Here $E = V/d$, $d = 20$ nm, the pore length. In Smoluchowski's theory of electrophoresis, the mobility is related to the surface (zeta) potential of the charged particle $\zeta$, $\mu = \varepsilon\zeta/\eta$, where $\varepsilon$ is the dieletric constant, $\eta$ the viscosity of the medium.

The best fit to the regime-II data in Fig.2(A) yields an apparent electrophoretic mobility value at $\mu = 1.95$ nm$^2$/μs-mV (or nm/μs per mV/nm), in convenient units. This mobility value is about 10x smaller than what is expected [22,23]. In addition to the effects of configurational entropy of the DNA which was attributed for the weak length-dependence [10] of translocation velocity, the electro-osmotic flow (EOF) $v_{EOF} = \mu_{wall}E$ which runs in the opposite direction of the electrophoresis, should also reduce the apparent mobility. The apparent zeta potential $\zeta$ should be the difference between those of the DNA and the negatively charged surfaces of the pore, i.e. $\zeta = \zeta_{DNA} - \zeta_{wall}$.[24,25] The effect of the EOF field is also implicated in the measured diffusion constant below.

As shown in Fig.2(B), in the low-voltage regime-I, the voltage dependence of D is wildly erratic. In contrast, in regime-II of high voltages, there appears to be a well-defined quadratic dependence on the applied voltage. We argue below that the quadratic voltage dependence of measured diffusion constant D can be understood as a consequence of a Taylor-dispersion effect due to the EOF flow field.

In pressure-driven flow in long cylindrical pipes, Taylor first showed [26,27] that the combination of radial diffusion and parabolic flow profile leads to a much enhanced effective diffusion constant along the longitudinal direction. The apparent longitudinal diffusion constant D is enhanced from the intrinsic diffusion constant $D_0$ by an amount $\delta D = r^2 v_0^2 / 192 D_0$ where r is the radius of the pipe, $v_0$ is the flow velocity at the center.

Here, in solid-state nanopores, the EOF flow field is expected to be spatially inhomogeneous, zero on the wall (no slip), maximum at distance ~ Debye length (~ nm) away from the wall. The EOF flow velocity profile should have a minimum at the center since over pore length of 20 nm the flow of positive counter ions cannot fully drag the rest of the fluid. For such an EOF flow profile, in Fig.3, during DNA translocation, the fluctuations in DNA's radial position lead to dispersions in the translocation velocity along the longitudinal direction. This dispersion effect will broaden the distribution of translocation time durations which in turn will lead to an apparent diffusion constant much larger than the intrinsic value. The change in D, $\delta D$, will be a quadratic function of the maximum velocity $v_{EOF}$ in the channel, since $v_{EOF} = \varepsilon \zeta_{wall} E / \eta = (\varepsilon \zeta_{wall} / \eta d) V$, thus $\delta D \sim r^2 v_{EOF}^2 / D_0 \sim V^2$, even though the numerical factor in the denominator will be different.

Nevertheless, one concrete result from this analysis is that by fitting the regime-II data in Fig.2(B) to $D = D_0 + aV^2$, where $a = r^2 (\varepsilon \zeta_{wall} / \eta d)^2 / D_0$ up to an unknown numerical factor, one obtains $D_0 = 11.2$ nm$^2$/μs. To compare this value with the Stokes-Einstein theory of diffusion, we model the translocating section of DNA inside the pore as a rod of length h, thus the intrinsic Stokes-Einstein diffusion constant $D_0 = k_B T / \gamma$, $\gamma = 2\pi \eta h / \ln(h/d_{DNA})$. By assuming h = 10 nm (half of the pore thickness d = 20 nm), and using $\eta$ = 1.8 cP (at 20% glycerol), dsDNA diameter $d_{DNA}$ = 2.2 nm, T = 300 K, we obtain $D_0 = 11.7$ nm$^2$/μs, in excellent agreement with the experimental value. The assumption here that the effective diffusing segment has a length of half the length of the pore is reasonable due to the hourglass shape of the inner part of the nanopores [28].

**Implications for nanopore DNA sequencing:**

With the finding of an analytical form Eq.(6) for the translocation times, which is at least of partial utilities at high voltages, one might ask what is the implication of Eq. (6) for determining spatial features (such as using hybridization probes [13] for sequencing DNA) on a DNA or other biological molecules.

Firstly, Eq. (6) means that the mean first-passage time is simply:

$$<\tau> = \int_0^\infty t F_1(t) dt = \frac{L}{v} \qquad (8)$$

In terms of spatial resolving power for using the nanopore technique, one needs to consider the mean-squared-first-passage time $<\tau^2>$ (the second moment) which is:

$$<\tau^2> = \int_0^\infty t^2 F_1(t) dt = \left(\frac{L}{v}\right)^2 + \frac{2DL}{v^3} \qquad (9)$$

Since a change in position will lead to a change in mean-first-passage time by an amount $\delta<\tau> = \delta L/v$, the necessary condition for resolving a spatial feature $\delta L$ away from position x = L would be

$$\delta<\tau> \geq \sqrt{<\tau^2> - <\tau>^2} \qquad (10)$$

or $\delta L/v \geq (2DL/v^3)^{1/2}$. Equivalently, this condition becomes

$$\delta L \geq \sqrt{2D \frac{L}{v}} \qquad (11)$$

Physically, Eq. (11) merely states that during the time of transit, L/v, for the nanopore to travel (in the reference frame of the DNA contour) from x=0 to x=L, the length scale of thermal smearing $(2DL/v)^{1/2}$ needs to be smaller than $\delta L$. Otherwise, this feature cannot be resolved. This condition is similar in principle to that in gel electrophoresis [29]. In the latter technique [29], when placed in an applied electric field inside a gel matrix, DNA fragments of different lengths have slightly different electrophoretic velocities, by an amount $\delta v$ (in gel matrix, the DNA electrophoretic mobility becomes length dependent). With increasing time, the separation between two bands grows as $\delta v t$. At the same time, the spread of each band grows as $(2Dt)^{1/2}$. The two bands become distinguishable when $\delta v t \geq (2Dt)^{1/2}$. Thus at the fundamental level, the nanopore technique and gel electrophoresis share the same set of basic physical principles.

**Summary**

We have re-examined the feasibility of using the first-passage time distribution based on a 1-D biased diffusion model to analyze DNA translocation data in nanopores in [1]. We showed that Schrödinger's first-passage probability density function gives an excellent description of DNA translocation time distribution in solid-state nanopores. At high voltages, we find that the extracted voltage dependence of the drift velocity is in excellent agreement with Smoluchowski's law of linear electrophoresis. Significant deviation, however, is observed at low voltages and is attributed to entropic barrier effects which undermine the validity of the model. At high voltages, the extracted diffusion constant is a quadratic function of applied voltage. We argue that such a behavior arises from the dispersive effect of electro-osmotic flow field inside the pore for the DNA translocation process. By assuming a Taylor-like dispersion form for the effective diffusion constant, we determine the intrinsic diffusion constant $D_0$ to be 11.2 nm$^2$/μs, a value in quantitative agreement with the prediction of Stokes-Einstein theory of diffusion. Finally, we urge the authors in [11,12] to carry out a re-analysis of their data using the correct FP-PDF formula here, as they used the incorrect formula from [1] as well.

**Acknowledgement:**
We are grateful to Prof. Jiali Li for her generosity in providing the data in [1] and for many helpful discussions. DYL wishes to thank the Siemens Foundation and Intel-STS for scholarships. XSL benefited from NIH-NHGRI grant R21HG004369 and useful conversations with Profs. Tom Chou (UCLA) and Tom Powers (Brown).

*All inquiries should be addressed to: xinsheng_ling@brown.edu

# Figure Captions:

**Figure 1**: Histogram of DNA translocation events from [1] and fittings using Eq. (6).

**Figure 2:** (A) The extracted drift velocity vs. applied voltage V. (Note: we assume contour length L=1040nm which is the sum of the contour length of 4 kbp DNA and 20 nm pore length.) The straight line is a fit of the regime II data to Smoluchowski's formula (see text). (B) The extracted diffusion constant D vs. applied voltage V. The line is a fit of the regime II data to $D = D_0 + aV^2$ (see text) with $D_0 = 11.2$ nm$^2$/μs.

**Figure 3**: A model for the effects of electro-osmotic flow (EOF) profile on DNA translocation. The DNA and the surface of the nanopore are negatively charged (not drawn). The expected EOF velocity profile (not to scale) for the center cross-section of the nanopore is shown by the red curve. The red arrow indicates the EOF flow direction, the black arrows shows the direction of translocation. Two positions of the DNA are shown relative to the EOF field profile for which the DNA sees different background EOF flow velocities.

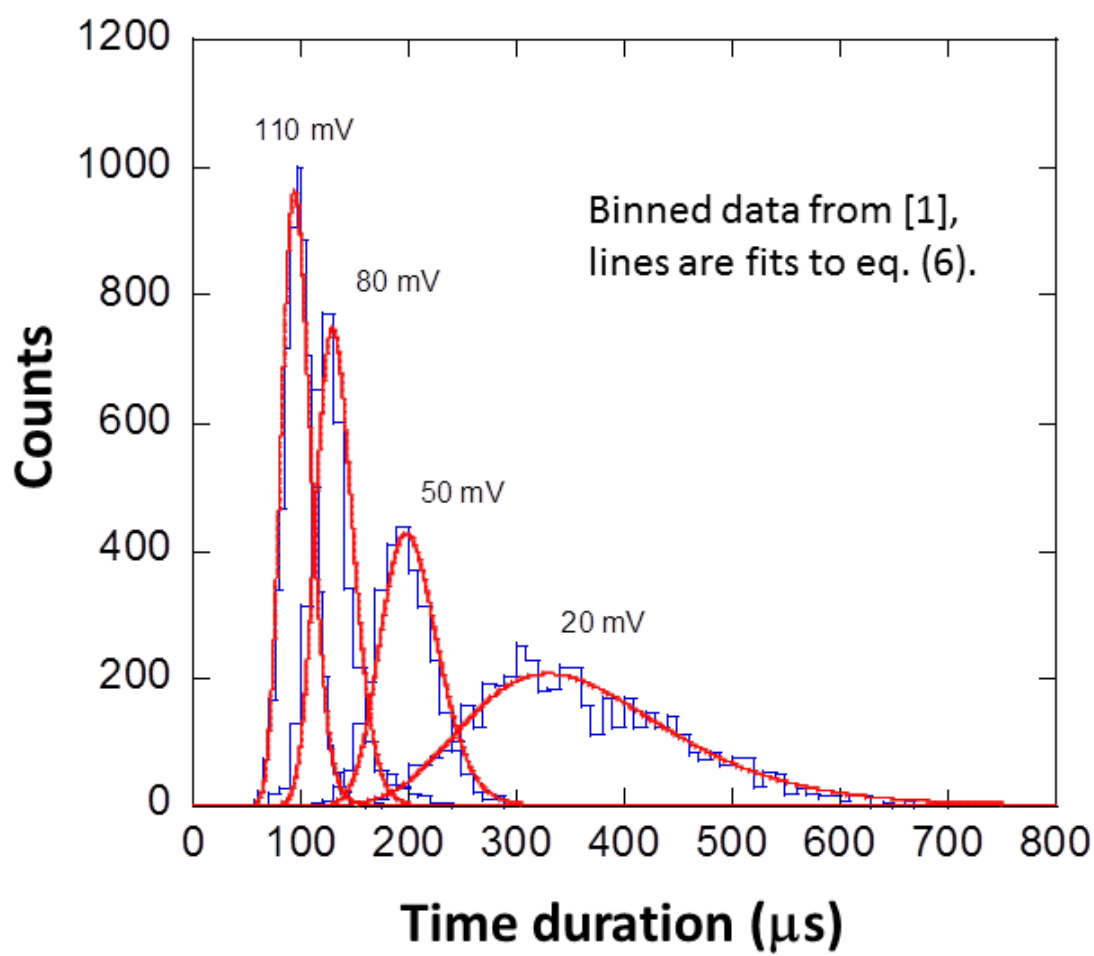

**Figure 1**

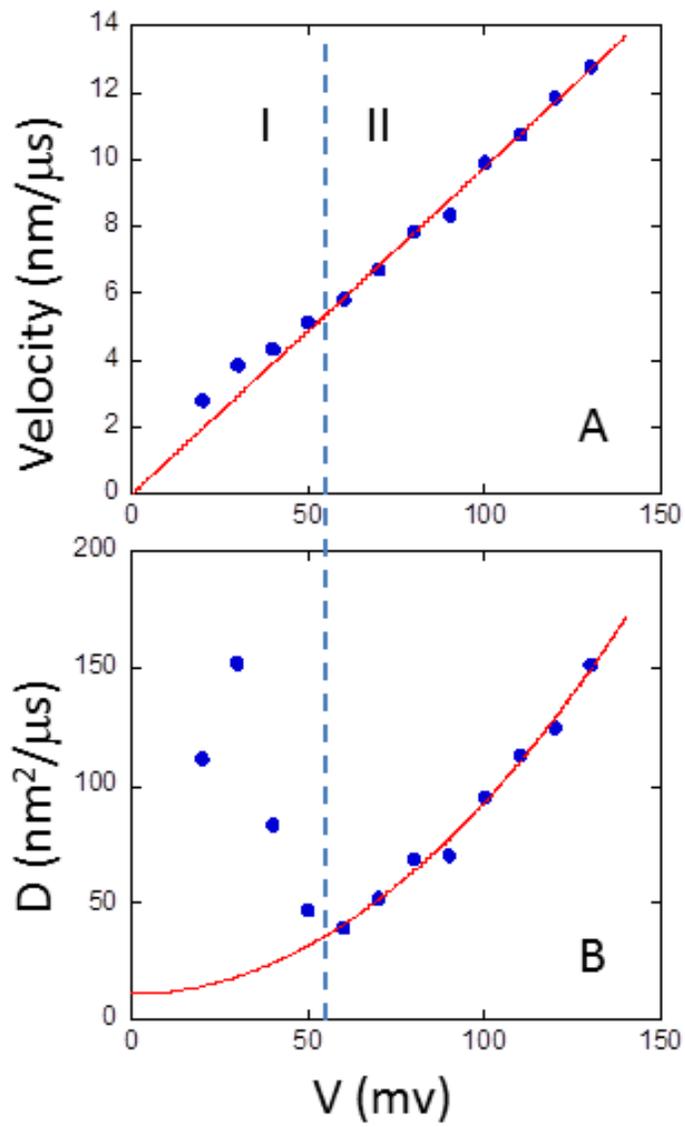

**Figure 2**

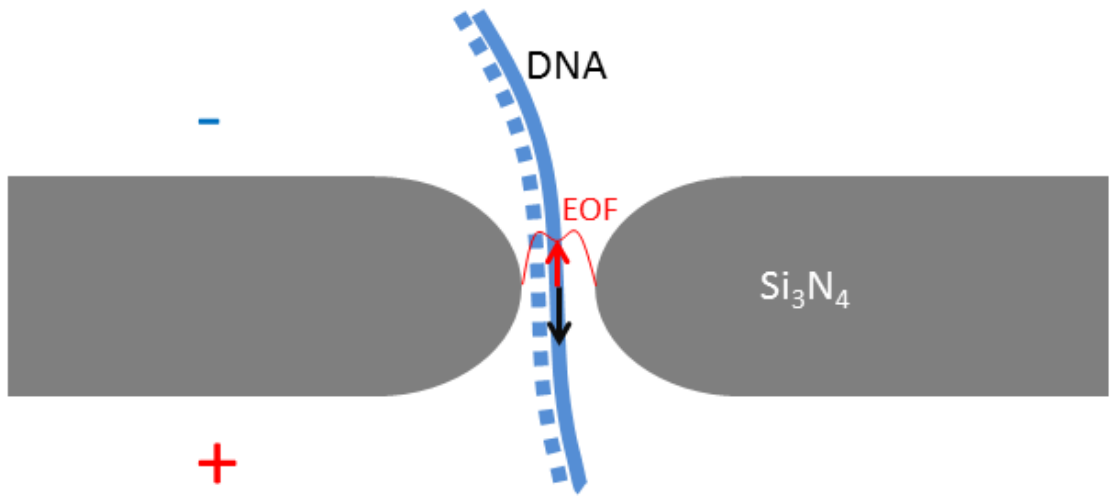

**Figure 3**